\pdfoutput=1
\documentclass[aps,prx,onecolumn,superscriptaddress,showpacs,floatfix]{revtex4-2}%

\usepackage[utf8]{inputenc}
\usepackage{amsmath}
\usepackage{amsfonts}
\usepackage{amssymb}
\usepackage{graphicx}
\usepackage{epsfig}
\usepackage{tabularx}
\usepackage{lineno}
\usepackage{setspace}
\usepackage[section]{placeins}
\begin{document}
\bibliographystyle{apsrev4-1}
\author{N.~Byrnes}
\affiliation{Department of Physics, University of Texas at Arlington, Arlington, TX 76019, USA}
\author{E.~Dey}
\affiliation{Department of Physics, University of Texas at Arlington, Arlington, TX 76019, USA}
\author{F.W.~Foss}
\affiliation{Department of Chemistry and Biochemistry, University of Texas at Arlington, Arlington, TX 76019, USA}
\author{B.J.P.~Jones}
\affiliation{Department of Physics, University of Texas at Arlington, Arlington, TX 76019, USA}
\author{R.~Madigan}
\affiliation{Department of Chemistry and Biochemistry, University of Texas at Arlington, Arlington, TX 76019, USA}
\author{A.~McDonald}
\affiliation{Department of Physics, University of Texas at Arlington, Arlington, TX 76019, USA}
\author{R.L.~Miller}
\affiliation{Department of Chemistry and Biochemistry, University of Texas at Arlington, Arlington, TX 76019, USA}
\author{L.R.~Norman}
\affiliation{Department of Physics, University of Texas at Arlington, Arlington, TX 76019, USA}
\author{K.E.~Navarro}
\affiliation{Department of Physics, University of Texas at Arlington, Arlington, TX 76019, USA}
\author{D.R.~Nygren}
\affiliation{Department of Physics, University of Texas at Arlington, Arlington, TX 76019, USA}
% ...
%

\title{Fluorescence Imaging of Individual Ions and Molecules in Pressurized Noble Gases for Barium Tagging in $^{136}$Xe}

\begin{abstract}The imaging of individual Ba$^{2+}$ ions in high pressure xenon gas is one possible way to attain background-free sensitivity to neutrinoless double beta decay and hence establish the Majorana nature of the neutrino.   In this paper we  demonstrate selective single Ba$^{2+}$ ion imaging inside a high-pressure xenon gas environment.  Ba$^{2+}$ ions chelated with molecular chemosensors are resolved at the gas-solid interface using a diffraction-limited imaging system with scan area of 1$\times$1~cm$^2$ located inside 10~bar of xenon gas. This new form of microscopy represents an important enabling step in the development of barium tagging for neutrinoless double beta decay searches in $^{136}$Xe, as well as a new tool for studying the photophysics of fluorescent molecules and chemosensors at the solid-gas interface.
\end{abstract}


\maketitle

The only known sensitive way to establish the Majorana
nature of the neutrino is via direct observation of neutrinoless double
beta decay ($0\nu\beta\beta$). Its detection would provide a revolutionary insight
into the nature of neutrino mass, likely the only observed manifestation
of physics above the electroweak scale. It would also be compelling
as a potential window into the mechanism leading to the dominance of matter
over antimatter in the Universe~\cite{buchmuller2005leptogenesis}.

All existing techniques to search for $0\nu\beta\beta$ have been
limited by backgrounds from radiogenic activity in detector materials~\cite{dolinski2019neutrinoless}.
To achieve sensitivity sufficient to discover $0\nu\beta\beta$ if the neutrino mass ordering is inverted and a light Majorana mediates the decay, detectors with $\geq$ 1 Ton of the double beta decay isotope and background
indices of order $b<0.1(\mathrm{ct\,ton\,ROI})^{-1}$ are required~\cite{agostini2017discovery}.
Factors between 20 and 2000 beyond existing demonstrated technologies
are required to meet these goals for the immediately forthcoming generation
of experiments. Beyond this ``ton-scale'' phase~\cite{albert2018sensitivity,adams2021sensitivity,abgrall2021legend,armatol2022toward}, the task of extending sensitivities into the normal mass ordering band of parameter
space appear truly formidable. The relevant experiments would employ
far larger quantities of active isotope~\cite{avasthi2021kiloton} and require background indices below the ton-scale demands. A distinct but related question is how to confirm a signal
of $0\nu\beta\beta$ following a suggestive but inconclusive hint
from the coming generation of still-background-limited experiments.
New and possibly radically new ultra-low background technologies will
be required to meet these challenges.

One especially promising technical approach to reaching the ultra-low
background limit is ``barium tagging''---
identification of the daughter ion produced in the double beta decay
of $^{136}$Xe~\cite{Moe:1991ik}. An efficient barium ion tag with low background could
reduce contamination from radiogenic backgrounds to near zero. 
Demonstration of a method of capture and imaging of barium ions from one to several tons of xenon requires significant advances in instrumentation.  Much progress has been made on promising methods for single barium ion or atom identification in liquid and gaseous xenon~\cite{Chambers:2018srx,mong:2014iya,rollin:2011gla,sinclair:2011zz,flatt:2007aa,mcdonald2018demonstration,rivilla2020fluorescent,jones2016single,thapa2021demonstration,thapa2020barium,Bainglass:2018odn,jones2022dynamics,herrero2022ba}, but many technological challenges remain.

Most approaches to barium tagging apply some form of fluorescence imaging, exciting  transitions in either atoms, molecules, or materials that are caused by presence of a barium atom or ion.   In xenon gas experiments, the target charge
state for barium tagging is the dication $\mathrm{Ba}^{2+}$, due to absence of recombination from thermalized electrons around the decay~\cite{novella2018measurement}. This ion
has no low-lying atomic fluorescence lines to access with visible
lasers. However, when chelated within a suitable organic molecule, its appearance can in principle be observed via single molecule fluorescence imaging
(SMFI). The proposal to use SMFI to tag the daughter ion in $0\nu\beta\beta$
was first outlined in Refs. ~\cite{nygren2016detection,jones2016single}. Commercial dyes developed
for $\mathrm{Ca^{2+}}$ sensing in biological applications were demonstrated
as sensitive barium tagging agents in solution phase. Soon thereafter,
single ion sensitivity was achieved~\cite{mcdonald2018demonstration}.  The model system in that work,
liquid droplets within a polymer matrix, is not well representative
of conditions within a xenon gas time projection chamber. Several aspects of the
chemistry of binding and fluorescence of the commercial dyes were
found to be inadequate for dry barium sensing, instigating a program
of novel organic fluorophore development, which has culminated
in multiple candidates for single $\mathrm{Ba}^{2+}$ sensing based
on crown-ether derivatives~\cite{thapa2019barium,thapa2020barium,byrnes2019barium,byrnes2019progress,rivilla2020fluorescent,herrero2022ba,miller2023barium}. These molecules have been used to
demonstrate single molecule sensitivity to barium in dry and solvent-less
conditions~\cite{thapa2021demonstration,miller2023barium} and shown via electron microscopy to react with neutral barium perchlorate in vacuum~\cite{herrero2022ba}. An active ongoing R\&D program is underway within the NEXT collaboration to bring these techniques to fruition in barium tagging sensors for a future ton- to multi-ton neutrinoless double beta decay experiment.

Experiments searching for $0\nu\beta\beta$ in $^{136}$Xe
typically employ time projection chambers filled with purified, pressurized or liquified
xenon at part-per-billion purity in oxygen and water to avoid electron
attachment.  To augment such a detector with a barium tagging system, two distinct problems must be overcome: first, how to bring the ion to a sensor~\cite{Bainglass:2018odn,jones2022dynamics} or sensor to an ion~\cite{Rivilla:2019vzd};  and second, how to sense the arrival of Ba$^{2+}$ with single-ion precision within a large, pure volume of xenon.  This is a complex environment in which to realize single ion microscopy, with no commercial devices or past proofs-of-principle available.

\section{Fluorescence imaging of single ions in high pressure xenon}

We present here novel results from NEXT barium tagging program that show a sensing scheme meeting many of the requirements
of single ion detection for barium tagging in $^{136}$Xe gas. In addition to this motivating application, this technology also opens a new field of single molecule / single ion imaging at the gas-solid interface.

\begin{figure}[t]
\begin{centering}
\includegraphics[width=0.6\columnwidth]{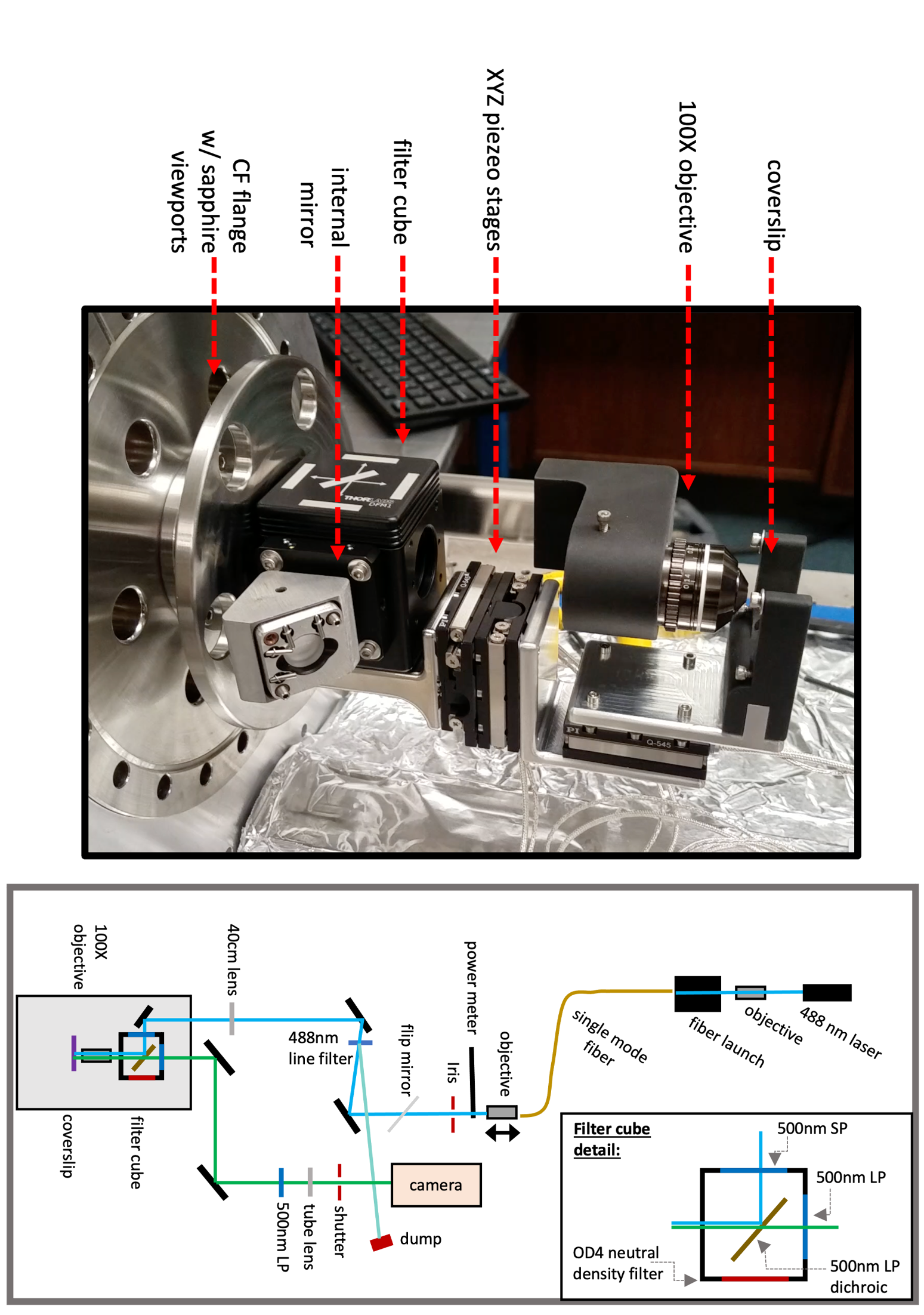}
\par\end{centering}
\caption{Top: the head of the high pressure microscope that sits inside the pressure chamber.r Bottom: Optical paths and components in the high pressure single molecule
microscope.\label{fig:OpticsImage}}
\end{figure}

\begin{figure}[t]
\begin{centering}
\includegraphics[width=0.99\columnwidth]{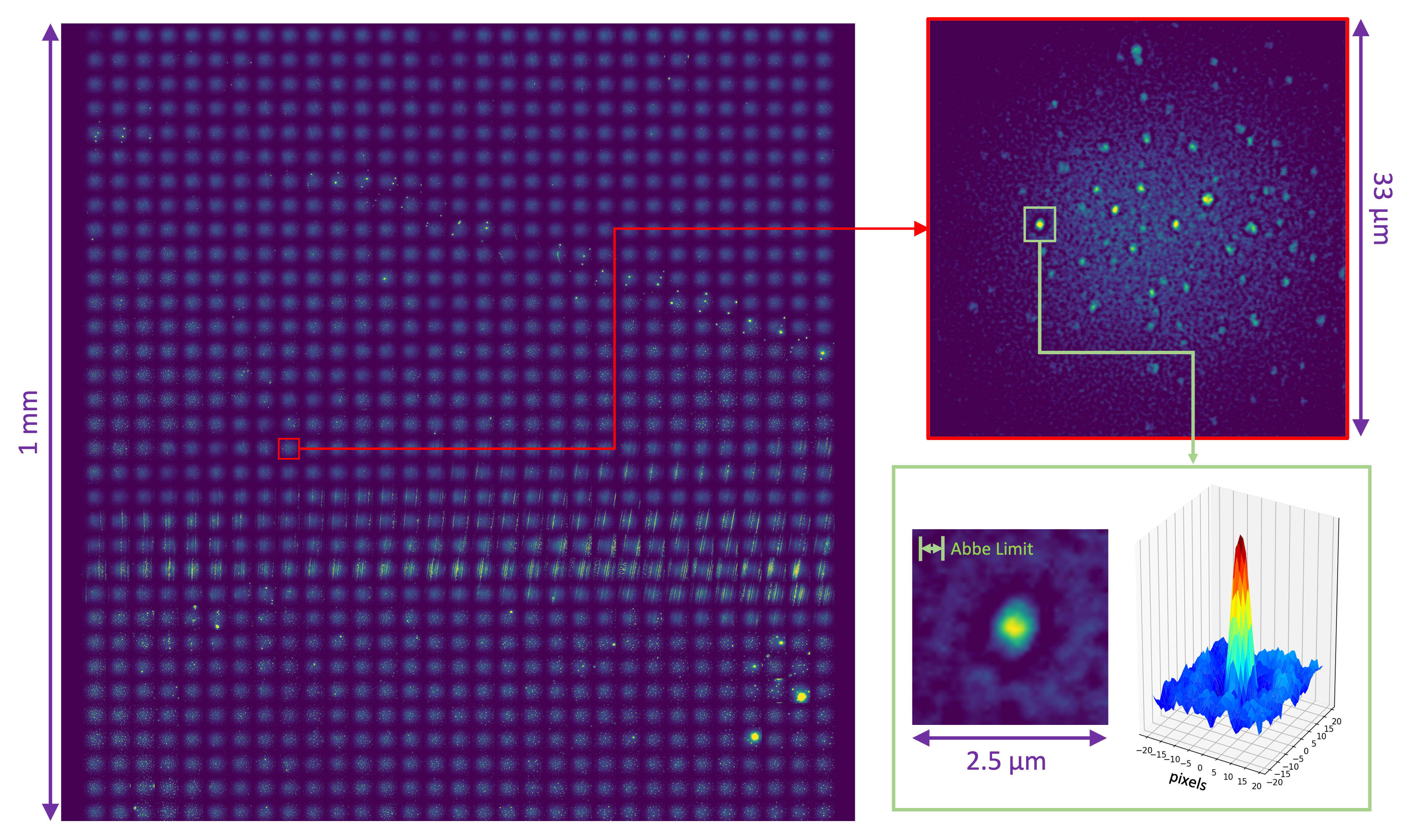}
\par\end{centering}
\caption{Large scale raw data image of BODIPY molecules drip-coated onto a slide surface. The image is resolved with point-spread function close to the Abbe
Diffraction Limit.  Bright points consistent with single molecule fluorescence are observed over a scan region of 1 mm$^{2}$.
\label{fig:Large-scale-image}}
\end{figure}

\begin{figure}[t]
\begin{centering}
\includegraphics[width=0.99\columnwidth]{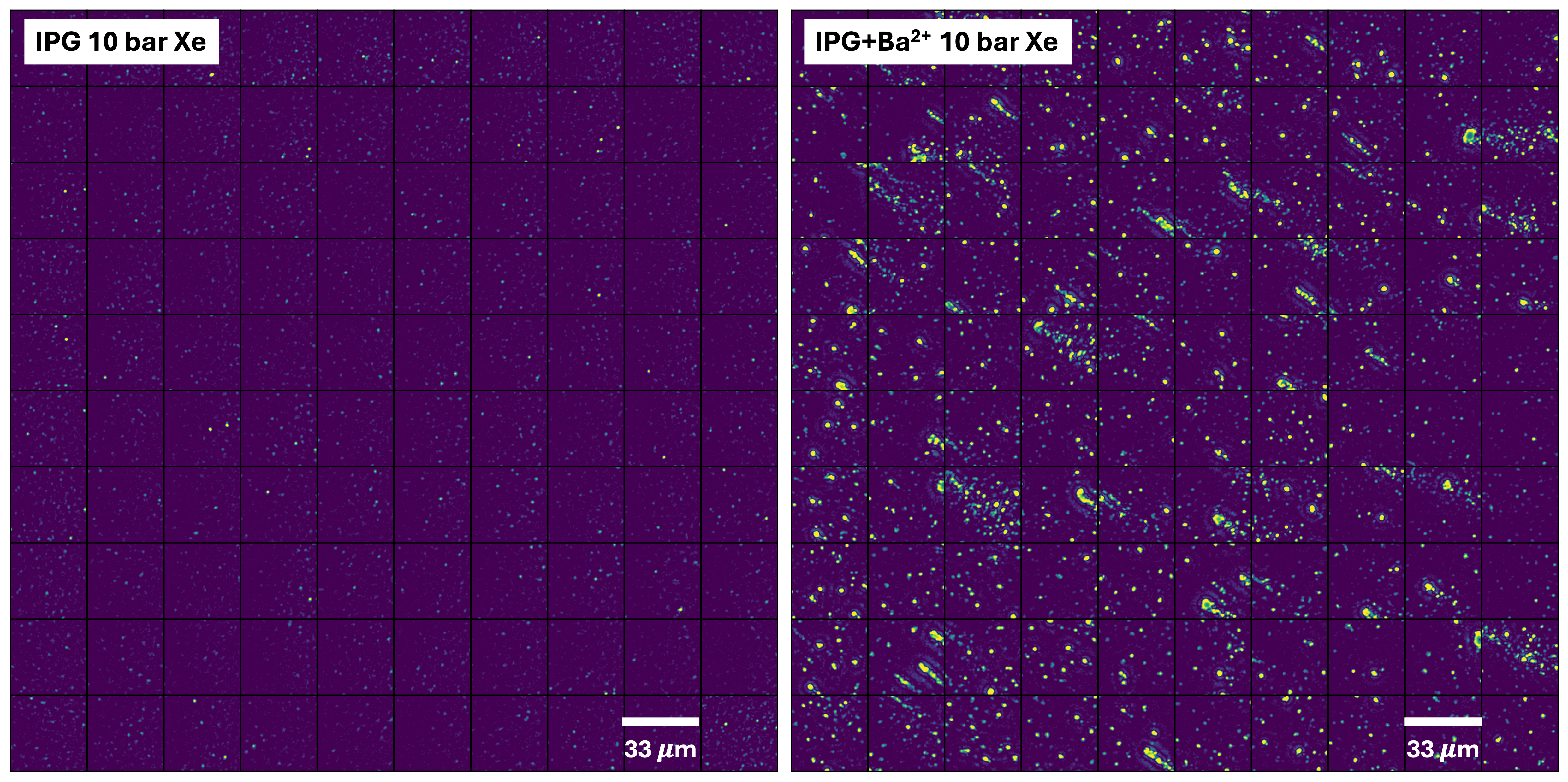}
\par\end{centering}
\caption{Image of a slide spin-coated in IPG-1 chemosensor, showing the activity with (right) and without (left) added Ba$^{2+}$. \label{fig:XenonImages}}
\end{figure}

\begin{figure}
\begin{centering}
\includegraphics[width=0.99\columnwidth]{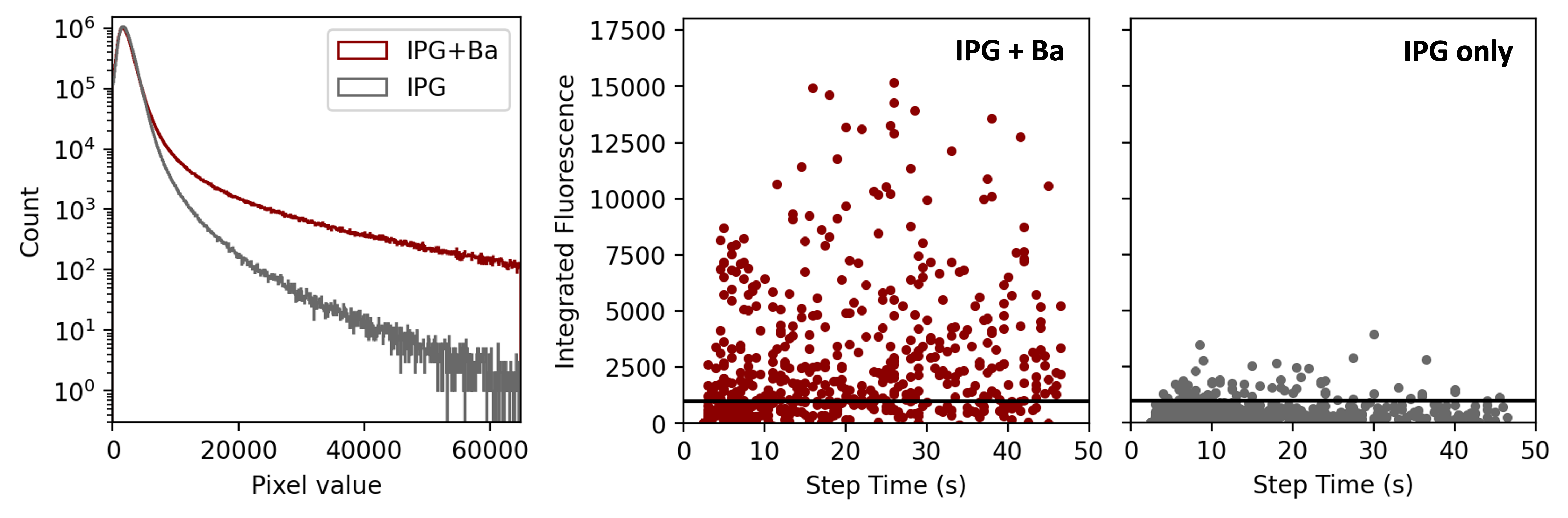}
\par\end{centering}
\caption{Left: Pixel intensity distributions for images taken using IPG vs IPG+Ba in 10~bar xenon gas. Right: reconstructed single molecule candidate brightnesses and step times for IPG vs IPG+Ba in 10~bar xenon gas. The plotted points are obtained over seven exposure regions on a single slide, and the trend is found to be repeatable over multiple slides. \label{fig:BaOnOff}}
\end{figure}

A detailed description of the instrument is provided in the Supplementary Information. Briefly, a fluorescence microscope based on nanometer precision vacuum stages is mounted on the inside of a high pressure chamber connected to a recirculating xenon gas system.  Images are recorded by an electron multiplying CCD camera, following laser excitation of barium-induced fluorescence in Ba$^{2+}$ selective organic chemosensors.   The system has been characterized using emission from BODIPY fluorophores~\cite{boens2012fluorescent} excited at 488~nm.  The configuration for this characterization study is shown in Fig.~\ref{fig:OpticsImage}.  A raw data image of sparsely distributed molecules, produced by rastering in 33~$\mu m$ steps over a 1 mm$^2$ surface area and auto-focusing on the single molecule candidates at each point is shown in Fig.~\ref{fig:Large-scale-image}.  The capability for single molecule imaging over this large surface area is clearly demonstrated, with resolution near the Abbe diffraction limit.  Further studies of the system imaging resolution can be found in the SI.

To demonstrate single Ba$^{2+}$ ion imaging in high pressure gas, slides were coated with nanomolar concentrations of IPG-1 fluorophore reported in Ref~\cite{miller2023barium}.  Solutions were prepared at 10$^{-8}$M concentration, with and without barium ions supplied via 100~mM barium perchlorate, and activity compared between Ba$^{2+}$ chelated and unchelated samples. The optical system for this test was configured with a 510~nm laser with 10~mW incoming laser power and a filter set suitable for this dye, as described in the SI.  The vessel was then sealed, evacuated and filled with
cleaned, pressurized xenon.  Figure~\ref{fig:XenonImages} shows images taken under purified 10~bar xenon gas for slides prepared with and without added Ba$^{2+}$.  In this image each frame has been Fourier transformed
and spatially filtered. A low pass filter is used to remove the broad
background from residual glass fluorescence, and a high pass filter
to remove speckles associated with thermal CCD noise. The resultant images display clear bright spots consistent with single molecule emission.

A clear increase in activity is observed upon addition of Ba$^{2+}$. Some visible streaks in the distribution of single molecule candidates emanate in a radial direction and are a result of the spin-coating protocol.  Single molecules remain resolvable within the streaks, though our analysis methods preferentially select those that are isolated single spots.  The raw pixel histogram indicates a dramatic increase in activity in bright pixels, as shown in, Fig.~\ref{fig:BaOnOff}, left. A small number of very weak emitters are present in the Ba$^{2+}$-free samples, though the bright spots associated with Ba$^{2+}$-bound IPG molecules are unambiguously identified as being present only in the Ba$^{2+}$-chelated sample.

\begin{figure}[tbh]
\begin{centering}
\includegraphics[width=0.6\columnwidth]{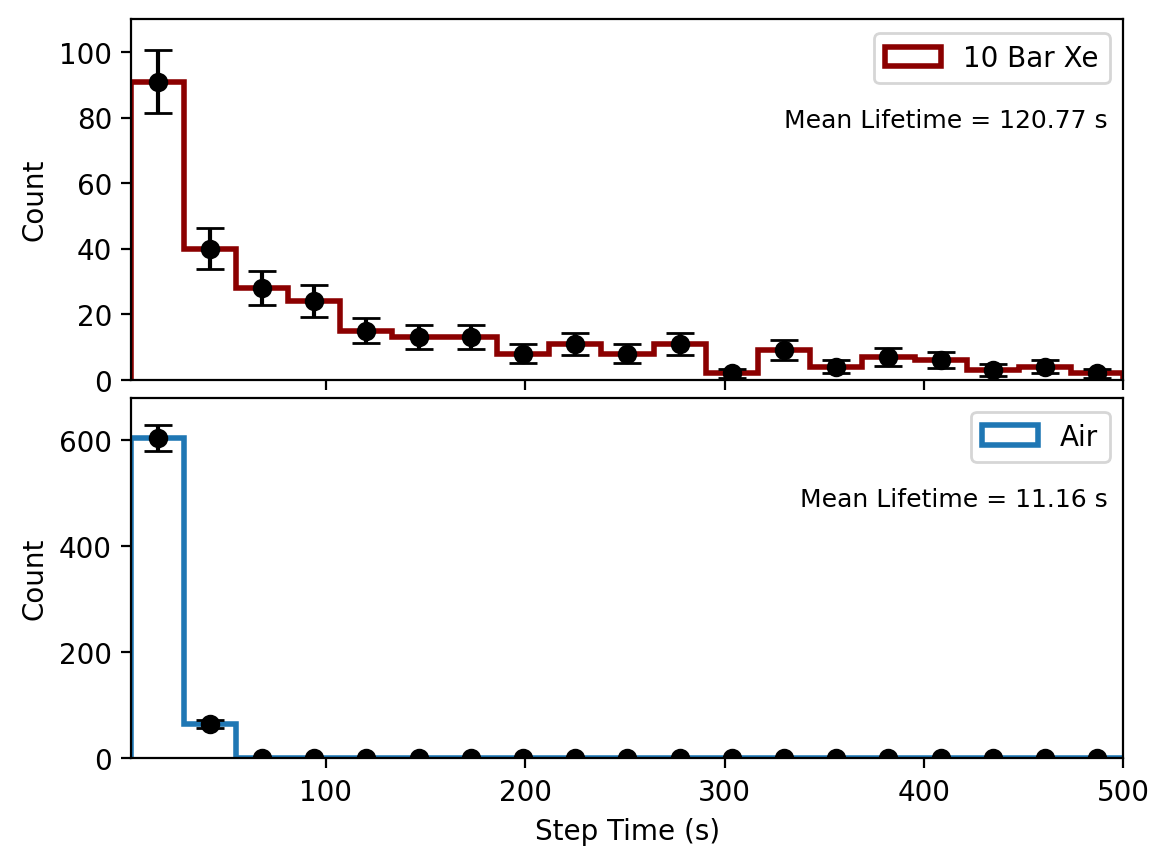}
\par\end{centering}
\caption{Photo-bleaching time distribution for Ba$^{2+}$ candidates chelated in IPG dyes in xenon and air. The X axis is shared between the top and bottom histogram.\label{fig:BaTimeConsts}}
\end{figure}

As a control experiment, blank coverslips with no fluorophore deposited were also scanned.  The characteristic diffuse glow from background fluorescence in the slide was observed in all locations. However, no localized bright emitters could be brought into focus at any point on the slides.  This suggests that the background due to accidental fluorescent molecules in the environment is extremely low, and has proven to be unquantifiably so in the present system. Because no emitters could be found, the microscope could not be focused and so no images from the blank slide surfaces can be reported.

The hallmark of single molecule fluorescent imaging is bright points that are characterized
by their sharp photobleaching transitions. A set
of analysis tools for single molecule fluorescence imaging and analysis
have been developed and are applied to the images captured by the
high pressure microscope.  To produce a time sequence, the first ten images
in a sequence are summed, and a list of local fluorescent maxima ordered by their intensity
is used to determine initial points of interest. Each frame is Fourier transformed and convolved with a bipolar filter kernel consisting of one normalized narrow Gaussian ($\pm4$ pixels) with positive
amplitude and one normalized broad Gaussian ($\pm8$ pixels) with negative amplitude. This filter serves to subtract local background while integrating fluorescence within the characteristic width of the fluorescent emission area.  A cut on the maximal identified step height in each trace is used to identify single molecule photo-bleaching event candidates.

Such candidates are identifiable in all conditions tested, including ambient air, vacuum and and pressurized argon and xenon, in Ba$^{2+}$ spiked samples.  The time and intensity distributions of the reconstructed steps in samples with and without Ba$^{2+}$ are provided in Fig.~\ref{fig:BaOnOff}, right.  In these plots, the $y$ axis corresponds to the fluorescence integral up to the photobleaching transition divided by the total time to the step, providing a measure of average brightness of the emitter. The $x$ axis corresponds to the step time.  Only very small steps are observed in the Ba$^{2+}$-free samples, whereas the Ba$^{2+}$-enriched ones contain large steps associated with single barium ion candidates.

\begin{figure}[tbh]
\begin{centering}
\includegraphics[width=0.85\columnwidth]{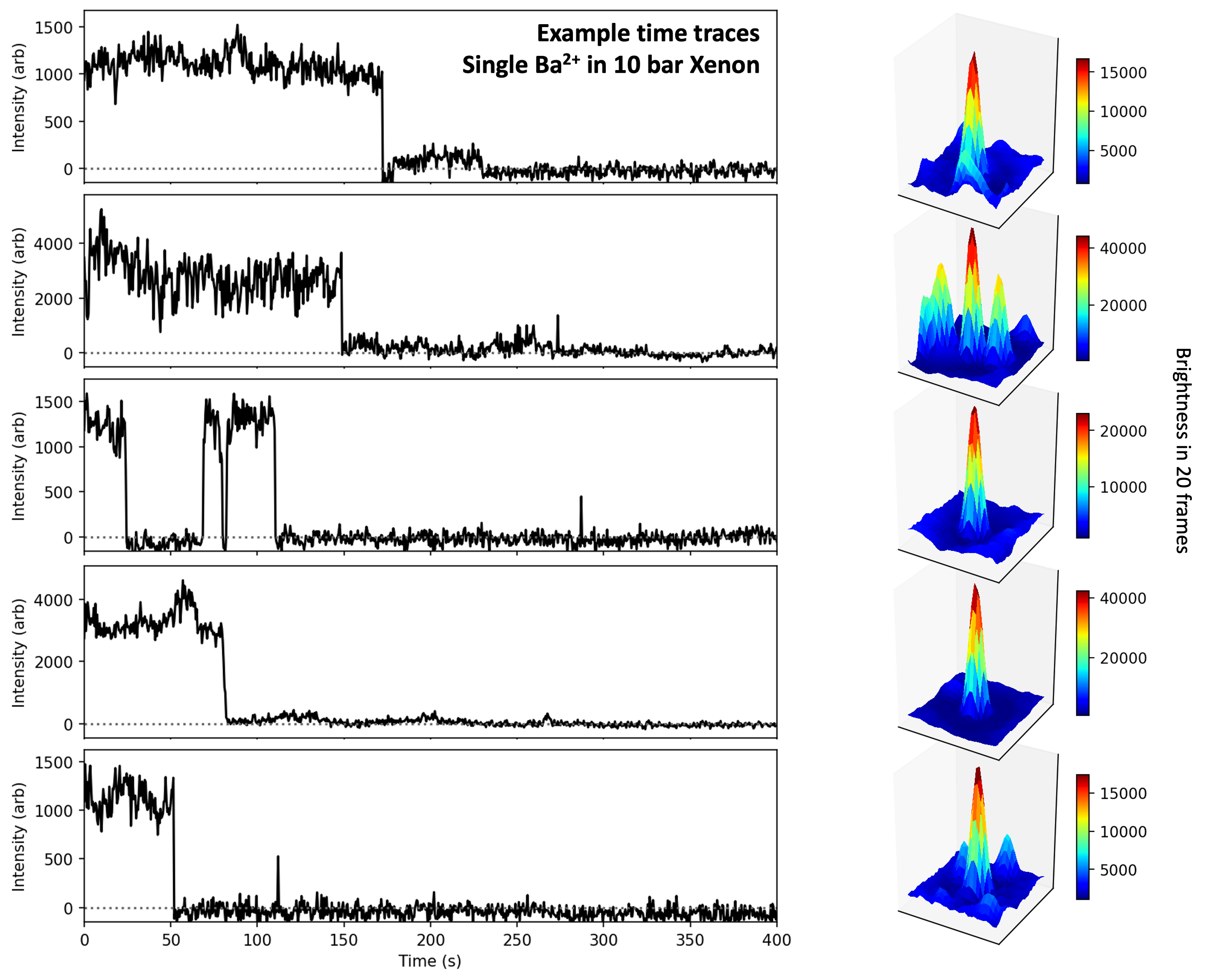}
\par\end{centering}
\caption{Single barium ions chelated with IPG-1 turn-on chemosensor
imaged in 10~bar xenon gas. The left plots show time traces of fluorescence for the identified emitter, with discrete photobleaching and photoblinking steps indicative of single molecule origin. The intensity is taken from the central pixel after applying a double-Gaussian filter as described in the text, and thus represents the baseline-subtracted integral over the diffraction-limited fluorescent spot. The right figures show the 2D spatial map of fluorescence around each emitter, integrated for 20 frames before photo-bleaching, showing a well localized peak in each case.  \label{fig:SingleBaXenon}}
\end{figure}

Fig.~\ref{fig:SingleBaXenon}, left shows the time traces for some single Ba$^{2+}$ ions identified via single-step photo-bleaching 10~bar xenon gas.   To the right of each trace are shown the fluorescence intensity maps corresponding to each, representing the activity recorded on a small subset of the CCD pixels.  Each surface plot shows the integral of the activity between in the last 20 frames before the photo-bleaching step.  The characteristic photo-bleaching and in some cases photoblinking behaviour associated with single molecule fluorescence imaging is clearly observed in each time series.

Studying the photo-bleaching time distribution illuminates a striking difference in the bleaching dynamics of Ba$^{2+}$-chelated IPG dyes in noble gas vs air environments, with substantially faster bleaching behaviour observed in air. The mean photo-bleaching time in 10~bar xenon gas is extracted as 241.5~s, whereas in air it is much faster at 22.3~s, as shown in Fig.~\ref{fig:BaTimeConsts}.  This observation supports the hypothesis that has been discussed in association with solution-based studies (for example, Refs.~\cite{zheng2014contribution,yokota2020fluorescence}), that reactions with oxygen are likely primarily responsible for the bleaching process.  Nevertheless, even absent ambient oxygen some photo-bleaching mechanism appears to be present, albeit at a far slower rate. While large datasets of IPG+Ba$^{2+}$ images were not collected in vacuum or argon conditions for this work, visual inspection of the photo-bleaching behaviour while imaging in those conditions showed a time profile far more similar to the the xenon data than to air, as may be expected under the oxygen-mediated hypothesis.

\section{Conclusions and outlook\label{sec:Conclusions}}

We have demonstrated a novel, diffraction-limited, high pressure fluorescence imaging system that is capable of single ion identification in high pressure gases at the gas-solid interface.  Single fluorescent molecules are resolved over large surface areas, and we have demonstrated a sweep over $1\times1$ mm$^2$ via 2D raster scan.  An autofocus algorithm that reliably brings emitters as weak as a single molecule into focus is used to provide a map of the focal plane that can be extrapolated for rapid and repeatable imaging.  The total possible scan region is in excess of $1\times1$ cm$^2$, with effective focusing possible over the full area.  

Single Ba$^{2+}$ ions have been imaged in high pressure xenon gas using turn-on fluorophores,  representing the first demonstration of imaging of individual Ba$^{2+}$  ions within a candidate active medium of a time projection chamber for $0\nu\beta\beta$.

The system can be implemented within the pressure vessel of a time projection chamber with only small modifications, which we now briefly discuss. First, in this work, rastering was performed by moving a slide
past a fixed objective lens, due to volume limitations within
the vessel.  In the final device it should be the converse, with rastering of the objective across a fixed imaging region. Since the image rays are brought out of the vessel in infinity space, this will add little in the way of complexity but requires some mechanical adjustments.  Second, continuous and lossless operation at this resolution generates a tremendous amount of data (15~GB uncompressed, for the image in Fig.~\ref{fig:Large-scale-image}). A rational zero suppression
algorithm and a form of online trigger for frames of interest is will be advantageous for real-time application, with factors of over 1000 in data reduction to be reasonably expected. These issues require attention, but appear manageable.  Finally, the molecular layer used here for Ba$^{2+}$ sensing is produced by spin coating a sparse group of molecules from solution; ultimately a fully sensitive Ba$^{2+}$ tagging layer must use densely packet fluorophores. Detailed investigations into the behaviour of densely packed fluorophore layers on surfaces~\cite{herrero2022ba} and into self-assembled monolayer growth~\cite{byrnes2019barium}, as well as studies of the efficiency of barium ion capture at these surfaces using Ba$^{2+}$ ion beams~\cite{navarro2023compact} are currently underway within the NEXT collaboration, and represent the next crucial frontier in chemical Ba$^{2+}$ ion sensor development. 
 While important work must be done to realize the ultimate sensing layers, an optical system like the one described in this paper will be capable of imaging ions arriving at their surfaces in a future xenon gas barium tagging detector.

Alongside ongoing development of molecular synthesis~\cite{thapa2019barium,thapa2020barium,byrnes2019barium,byrnes2019progress,rivilla2020fluorescent,herrero2022ba,miller2023barium}, ion transport~\cite{jones2022dynamics} and detector readout modalities~\cite{soleti2023towards,martins2024high,villalpando2020improving,byrnes2023next} that can enable barium tagging at the cathode, the new technology demonstrated in this paper represents an important step toward a large, barium tagging xenon gas detector. Such a device holds great promise as a concept for a truly background free, ton-to-multi-ton scale neutrinoless double beta decay experiment in $^{136}$Xe. It also represents a first step into a new field of high precision, single molecule photo-physical fluorescence analysis at the gas-solid interface.

\section*{Acknowledgements}

This program of single barium ion sensing is a part of a collaborative project to employ SMFI chemosensors functional at the solid-gas interface, as R\&D toward barium tagging for the NEXT experiment.  We acknowledge support from the US Department of Energy under awards DE-SC0019054 and  DE-SC0019223, and US National Science Foundation under award number NSF CHE 2004111 and the Robert A Welch Foundation under award number Y-2031-20200401 (University of Texas Arlington).  

The NEXT Collaboration acknowledges support from the following agencies and institutions: the European Research Council (ERC) under Grant Agreement No. 951281-BOLD; the European Union’s Framework Programme for Research and Innovation Horizon 2020 (2014–2020) under Grant Agreement No. 957202-HIDDEN; the MCIN/AEI of Spain and ERDF A way of making Europe under grants RTI2018-095979 and PID2021-125475NB , the Severo Ochoa Program grant CEX2018-000867-S and the Ram\'on y Cajal program grant RYC-2015-18820; the Generalitat Valenciana of Spain under grants PROMETEO/2021/087 and CIDEGENT/2019/049; the Department of Education of the Basque Government of Spain under the predoctoral training program non-doctoral research personnel; the Portuguese FCT under project UID/FIS/04559/2020 to fund the activities of LIBPhys-UC; the Pazy Foundation (Israel) under grants 877040 and 877041; the US Department of Energy under contracts number DE-AC02-06CH11357 (Argonne National Laboratory), DE-AC02-07CH11359 (Fermi National Accelerator Laboratory). Finally, we are grateful to the Laboratorio Subterr\'aneo de Canfranc for hosting and supporting the NEXT experiment.

\bibliography{microscope.bib}

\newpage

\section{Methods}

To obtain high quality images, any fluorescence microscope must have carefully arranged optical paths for both illumination and image collection.  We begin by describing these optical elements in Sec.~\ref{sec:Excitation} and ~\ref{sec:Emission}, respectively, and then describe the mechanical construction of the device in Sec.~\ref{sec:Mechanical}. Section~\ref{subsec:Focusing} describes the alignment and focusing protocol and quantifies some of the imaging metrics of the system.

\subsection{Excitation~\label{sec:Excitation}}

Past work on single Ba$^{2+}$ imaging for NEXT has shown that single ion candidates can be observed with at little as 0.2 mW of laser power in the field of view (FOV), corresponding to around 1~W cm$^{-2}$~\cite{mcdonald2018demonstration}, given molecules with sufficiently high quantum efficiencies. Laser power above a 50~W cm$^{-2}$, on the other hand, risks rapid photobleaching, preventing the prolonged observation needed to tag a single emitting molecule, for all of the molecular species we have studied. As such we have optimized the current system to provide few-mW levels of excitation to enable clear single molecule resolution, while offering suitably long (few second) observation times to establish presence of an individual ion.  We have also opted for an epi-fluorescence configuration.  While through-objective total internal reflection fluorescence (TIRF) imaging was used in our earliest studies of fluorescent single barium ion complexes in order to suppress deeper backgrounds arising in thick samples~\cite{mcdonald2018demonstration}, more recent work demonstrated that for sufficiently thin fluorescent samples, single molecule resolution can also be obtained in an epi-fluorescence mode~\cite{thapa2021demonstration}. The latter approach is more straightforward to realize with remote illumination, and forms the basis of our illumination scheme for this device. For this form of microscopy, light is focused on axis on the back-focal plane of the objective by an external lens, leading to parallel and uniform passage of light through the focal plane, which is aligned to the surface supporting the fluorescent emitters to be imaged

We use a series of solid state lasers as excitation sources, each suitable for a different set of fluorescent dyes.
For the characterizations of the optical system performance, we have used a 488~nm laser with power controlled by an adjustable DC power supply.
This wavelength choice is sufficiently long to escape the tail of fluorescence from glass that compromises single molecule imaging with shorter wavelength excitation (at 450~nm and below, the autoflourescence from the objective and substrate proved prohibitive for single molecule imaging), and is well matched to the absorption peak of the BODIPY~\cite{boens2012fluorescent} molecule used for our optical system characterizations.  For Ba$^{2+}$ sensing, a 510~nm laser is used to excite the IPG-1 species studied in Ref.~\cite{miller2023barium}.  Due to their long excitation wavelength, the IPG class of molecular probes have have been found to provide excellent signal-to-background ratio for single molecule microscopy, in studies undertaken in preparation for this work~\cite{miller2023barium}.  An illustrative diagram showing the system configured with the 488~nm laser is shown in Fig.~\ref{fig:OpticsImage}, top left.

To obtain smooth excitation profiles over the image plane, the excitation laser beam is first spatially cleaned. For the 488~nm laser, it is first
 launched from the laser into a single mode optical fiber~\footnote{Thorlabs P1-405B-FC-2} via a 20X microscope objective~\footnote{Olympus PLN20X with 0.4 NA}.  At the other end
of the fiber a 10X objective~\footnote{Olympus PLN10X} produces a parallel Gaussian beam of around 2~mm diameter.  Longitudinal positioning of this second objective on a micrometer stage allows for fine-tuning of the size of the illumination site on the
sensing plane by controlling the beam divergence from the objective.  For the 510~nm laser, the beam is instead launched into a 5x beam expander, with adjustable divergence, and then spatially filtered through a pinhole. 

Immediately down-beam, a flip-mirror
can be used to redirect the laser over a 4~m path to quantify its divergence. Optimal illumination performance was found to correspond to a very slightly
divergent beam from this mirror. The excitation light is then passed through an adjustable
iris to remove halo. The 488~nm laser configuration then includes a 488 nm laser line filter~\footnote{Thorlabs FL488-10}; whereas for the 510~nm configuration, no subsequent excitation filtering proved to be necessary external to the vessel. The
reflected ray from the line filter contains the long and short wavelength
tails of the laser spectrum and is absorbed on an external beam dump. 

The spatially and spectrally cleaned laser light is reflected from two
external adjustable mirrors in a periscope arrangement to allow for
fine steering of the path into the vacuum / pressure chamber. The beam next passes through a 1.6~mm thick sapphire pressure
window into the vacuum or gas volume.  Just in front of the sapphire window, a final 40~cm
spherical planoconvex lens is placed at approximately its focal length from the main microscope objective,
focusing the nearly parallel beam onto the back-focal plane. 

Inside the chamber the beam arrives at a fine-adjustment mirror with
two degrees of freedom, which is set before the chamber is closed
and left aligned for the duration of the experiment. This mirror directs
the beam into an internal filter cube, which has emission and excitation filters and a long-pass dichroic beam splitter.  For the 488~nm configuration, these are  500~nm short pass (SP) excitation~\footnote{Thorlabs FEHH0500} and 500~nm
long pass (LP) emission~\footnote{Thorlabs FELH0500} filters, and a 490~nm dichroic mirror, also LP~\footnote{Thorlabs DMLP490R}. For 510~nm, a  fluorsescence microscopy filter set~\footnote{Chroma 49023} consisting of excitation filter: band pass 500$\pm$ 20~nm, emission filter: band pass 560$\pm$ 25~nm, and dichroic mirror: long pass 525~nm.
On the forth side of the filter cube a high optical density neutral
density filter serves as a shallow-depth internal beam dump.  

The excitation
beam is guided to the back face of a vented 100X, high NA microscope
objective~\footnote{Customized PLFLN100X; PLAN FLUOR 100X DRY OBJ, NA 0.95, WD 0.2} which we have customized for operation in high pressure environments.  Earlier experiments with commercial objectives resulted in various bursting and internal misalignment failures due to the pressurization and depressurization process. 
The beam is focused by the objective onto the front face of a 160~$\mu$m thick glass coverslip~\footnote{Ted Pella Schott D263M 22x30~mm Glass Coverslips}
in transmission mode, and the fluorescence emitted from the sample is
collected back through the same high numerical aperture objective. 

\subsection{Imaging~\label{sec:Emission}}

The longer wavelength fluorescence light emitted from the sample plane transmits through
the dichroic mirror and exits through a second 2-inch diameter sapphire viewport~\footnote{CeramTec 18617-01-CF}.
The infinity-space optical path on the fluorescence side is completely enclosed in
black optical piping to prevent ambient background from laboratory lighting.  The light
isolation in the image path is found to be sufficiently effective that single molecule resolution is
comfortably accomplished with laboratory lights on. The image is
reflected twice from a pair of mirrors in a periscope arrangement
that allow for external adjustment of the region of the objective
in the camera FOV. A second, external 500~nm LP filter~\footnote{Thorlabs FELH0500} removes
any residual short-wavelength light, and then a tube lens focuses the image from
infinity space onto the camera. Adjusting the tube lens position
and focal length allows for the system to be run in modes with different
levels of magnification. For this paper we have used a medium level of magnification, with approximately 40 $\mu$m field of view per frame 
achieved using a 15~cm focal length tube lens. 

The image acquisition device is an electron multiplying CCD camera~\footnote{Hamamatsu ImagEM2 EMCCD} with 90\% quantum efficiency
and a 512$\times$512 array of $16~\mathrm{\mu m} \times 16~\mathrm{\mu m}$ pixels (pix), operating at 500~ms exposure time.  The camera is connected
via a demountable tube coupling with an external shutter to protect
the camera from being exposed to high light environments.  It is read out over IEEE 1394 conduits to a PC running a bespoke data acquisition software suite that we have developed specifically for this device that interfaces to the internal micrometer stages and camera readout.

\subsection{Alignment and Focusing~\label{subsec:Focusing}}

\begin{figure}
\begin{centering}
\includegraphics[width=0.9\columnwidth]{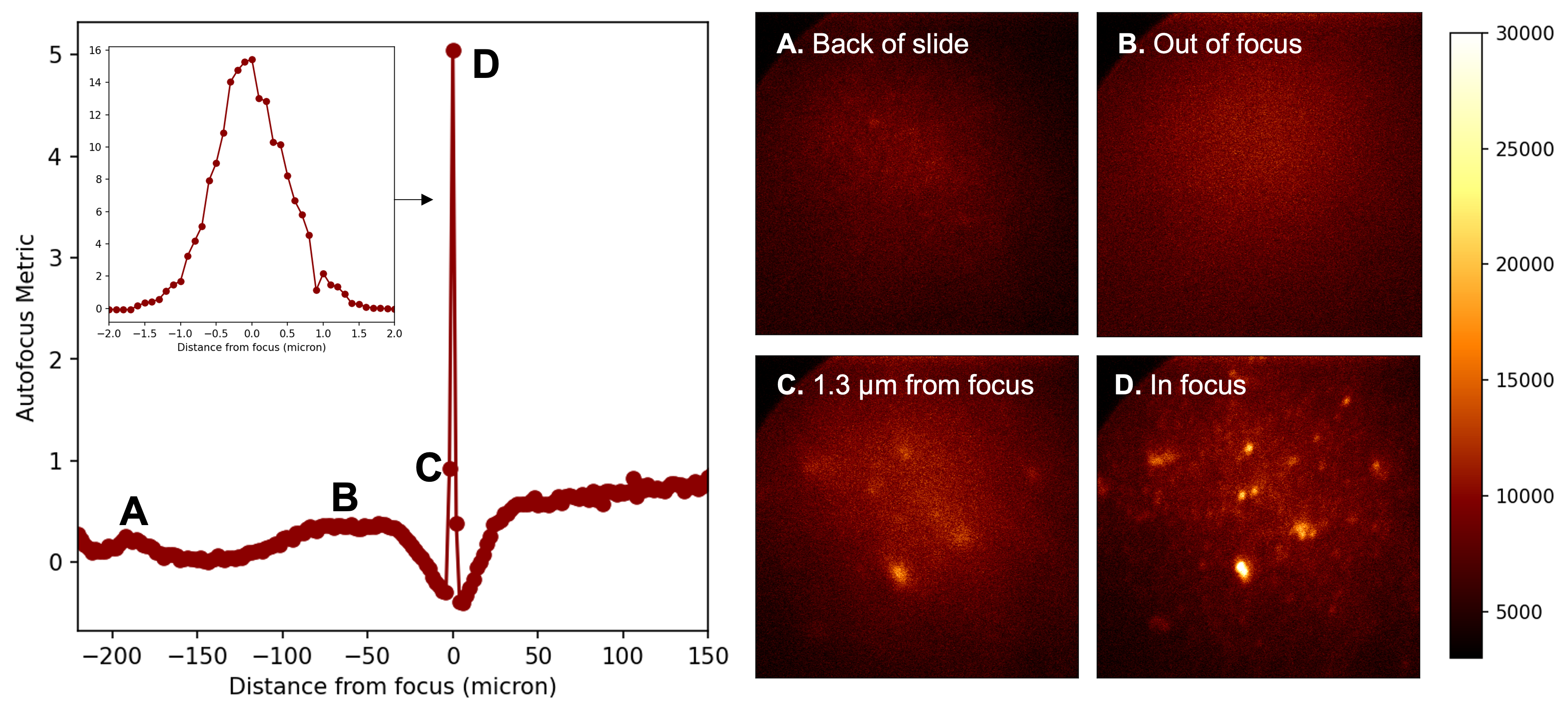}
\par\end{centering}
\caption{Illustration of the single molecule autofocus procedure. An autofocus
metric defined in terms of image brightness and kurtosis is applied
that maximizes for an in focus image. The left plot shows the Z dependence
of the autofocus metric, with images at different focal depths provided
on the right plot. A small signal is seen on the back face of the
slide (A). No features are present when totally out of focus (B).
Within around 2 microns of focus some activity is seen (C), with sharp
images only at the focal plane (D). The inset shows that the depth
of the in-focus region is around $1\mu m$.\label{fig:Autofocus}}
\end{figure}

\begin{figure}[t]
\begin{centering}
\includegraphics[width=0.9\columnwidth]{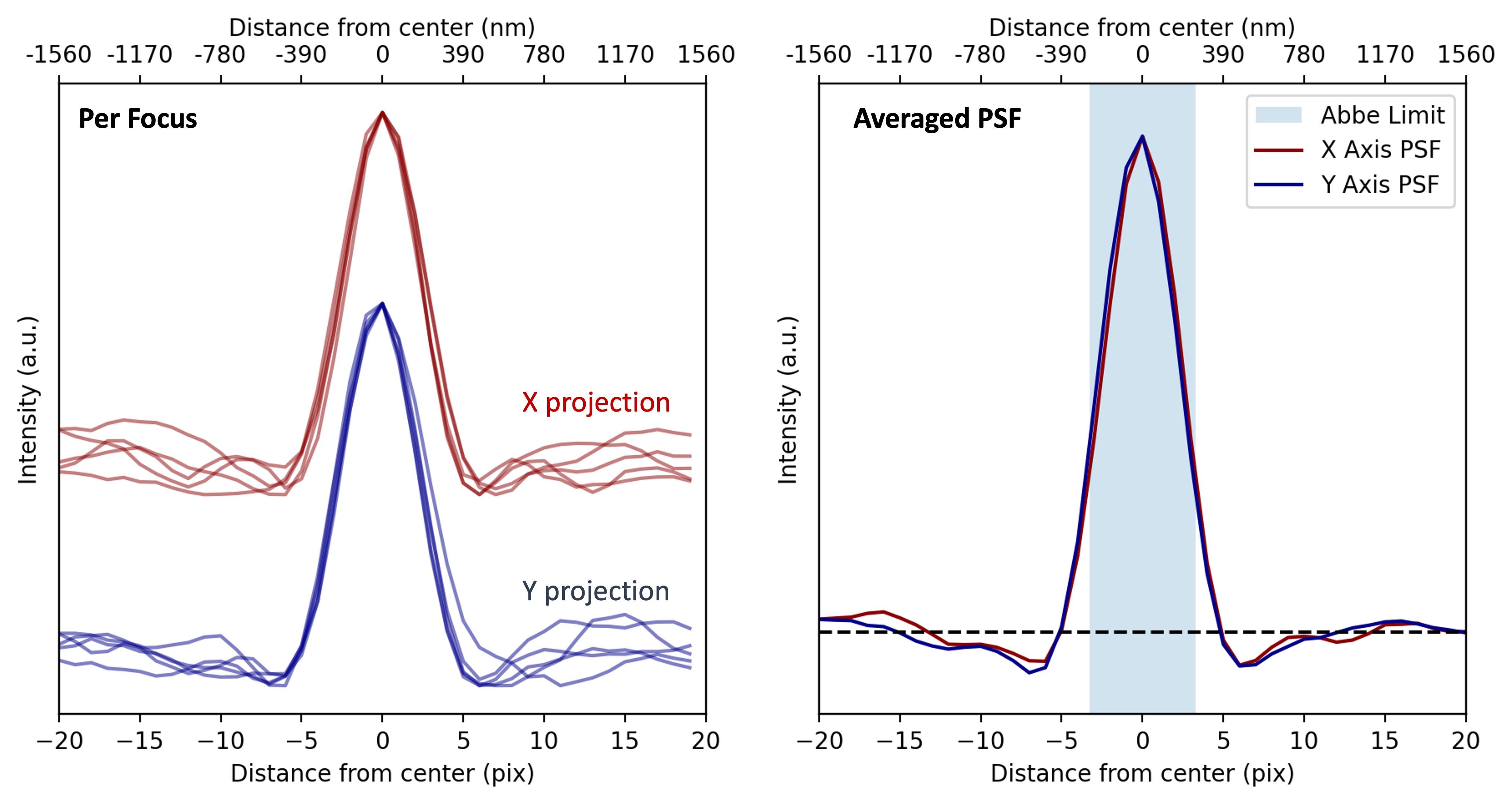}
\par\end{centering}
\caption{Left: Measured point-spread function in X and Y directions at five re-focused locations using single molecules in 7~bar argon gas. Right: average X and Y PSF compared to the Abbe diffraction limit. The horizontal line on the right plot represents the baseline, as measured away from the bright spot.
\label{fig:PSF}}
\end{figure}

At each $X$ and $Y$ position there is an approximately $1~\mu$m deep range of $Z$ values where the
fluorescence plane is in focus. Finding this plane at each position requires identification
of a weak single molecule fluorescence signal among the range of plausible focal points, which vary by around 70~$\mu$m in absolute terms given repeatability afforded by our slide
installation protocol. Few $\mu$m variations of the focal plane
depth with temperature are observed, consistent with expectations based on thermal expansion of structural materials and the range of temperatures recorded.  Furthermore, when changing between gaseous and vacuum conditions we observe a slow drift of the focal plane position on few-hour timescales, consistent with the expected effect of swelling of  plastics in the structure through gas absorption and degassing~\cite{rogers2018high}.   These effects taken together imply that a dynamical method of establishing and tracking the focal plane is required.

To find the focal plane,  a single-molecule-sensitive
autofocus mechanism has been developed. The focal metric is the ratio of the pixel-wise
kurtosis over the mean pixel intensity, which is maximized at each $X$,$Y$ to establish the focal plane location in stage travel coordinates.
This metric favors images with a small number of bright pixels among a field
of dimmer ones, which is the characteristic of an in-focus image. Occasionally a cosmic ray muon or radioactive event passes through the EMCCD
leading to one or a few extremely bright pixels, presenting to
the autofocus metric as an anomalously in-focus image. We exclude such
anomalous events by rejecting images with up to 5 bright pixels among
a field of otherwise dim ones. This signature can also be distinguished from that associated with an extremely bright fluorescence emitter by its lack of persistence when stationary in the fluorescence
plane.

Figure~\ref{fig:Autofocus} illustrates the performance of the autofocus metric using single BODIPY molecules. Far from the focal plane there are no discernible
features, either in the images or in the metric. A
set of very dim features are visible on the back of the slide. Since
there is no fluorescent layer coated onto this back surface,
we interpret these features as being associated with small amount of out-of-focus fluorescence light scattering from optical imperfections on the reverse of the glass. Inside the cover-slip volume no visible features
are present. On the far side of the coverslip where we approach the desired
focal plane, a sharp spike in autofocus metric is apparent. A higher
resolution scan in this region (inset in Fig~\ref{fig:Autofocus}, left) shows
that the depth of focus is around 1 $\mu m$. Maximizing the autofocus metric is reliably found to obtain focal plane to $\pm0.3$~$\mu$m precision.

After focusing, the images of single point-like emitters exhibit consistent point-spread function (PSF) in air, vacuum and pressurized gases.  Fig.~\ref{fig:PSF} shows the $X$ and $Y$ projections of the measured point-spread function obtained using single  molecules within 7~bar of argon gas.  These PSF projections are obtained by averaging over the the $X$ and $Y$ directions around the brightest 20 fluorescent emitters, re-focusing in 5 distinct locations (Fig.~\ref{fig:PSF},~left).  The width of the PSF depends slightly upon how perfectly the focal plane has been obtained, but in all cases is very similar to the expectation from Abbe's theory of diffracting optics~\cite{born1997principles}. The averaged PSFs in the $X$ and $Y$ directions are compared directly to the Abbe limit in Fig.~\ref{fig:PSF},~right.  These results suggests that the microscope essentially saturates the theoretical limit of optical resolution, even when running within a pressurized noble gas environment.  We note that, due to its single molecule sensitivity, super-resolution techniques can also be used in this system to advance beyond the diffraction limit, though this is of limited value for the barium tagging application.

Once a slide is installed, the standard procedure is to find the focal $Z$ position at 5--10 points on the face of the
slide. These data are then used to generate a 3D map of the focal plane by extrapolating a 2D surface through the data points. This
fitted focal plane can then be used to find the focus at any other
point for subsequent imaging.  The focal plane map continues to accumulate data points as in-focus images are found, which serves to continuously refine its precision for extrapolation to new points.  This approach allows the focus to be found rapidly at new imaging locations, and thus allows for construction of larger, rastered images by scanning over X-Y positions.

\subsection{Mechanical design and gas handling \label{sec:Mechanical}}

The high pressure microscope system is built into a 16 inch long, 6 inch diameter custom-manufactured
pressure vessel with 8-inch ConFlat flanges on both ends. The microscope front end-cap is fixed down
to an air-levitated optical table with a thick aluminum bracket. Vibration isolation was found to be crucial to achieving optimal resolution, and careful positioning of the various vacuum pumps and circulation pumps around the optical table proved to be instrumental in achieving sharp images.  The
pressure vessel pipe and back end cap slide on a set of two parallel
rails to open and close the vessel, leaving the microscope head fixed in place in order to maintain rough alignment when opening and closing the large CF flanges.    

Inside the vessel and cantilevered from the front end-cap, a machined aluminum frame supports a small HDPE bracket that holds the internal microscope objective in a fixed position.
A 3-axis vacuum stage system~\footnote{PI Q545.140} is used to maneuver and monitor a microscope slide in three dimensions in front of the objective. The stage is rated for ultra-high vacuum and has nanometer precision along all three axes, with an active feedback loop inside the device.  Our specifications
demand only few hundred nanometer precision, below which diffraction limits the point spread function, and the stage  comfortably meets these requirements.  Careful frequency tuning of the feedback mechanism within the stages had to be made in order to avoid exciting resonant mechanical normal modes of the cantilevered system that inhibited stable positioning.   The full range of motion of the stages is  $\pm6.5$ mm in each direction, though our control software  restricts the $X$ and $Y$ (in-plane) motion to a region of 10$\times$10~mm and the range of $Z$  (focusing) motion to within 1.5~mm in front of the microscope slide, to avoid damage from scratching the objective.

The system is evacuated using a turbo pumping station~\footnote{Pfeiffer HiCube 80 Eco} which can be decoupled
using an isolation valve~\footnote{Carten HB-51} when pressurized. Despite the use of some
plastics in the system, the vacuum quality as monitored by a hot filament ion
gauge~\footnote{Kurt Lesker KJLC 354} routinely reaches $10^{-6}$ Torr prior to filling with gas, which is an adequate vacuum quality for subsequent fill and operation of time projection chamber detectors.
Pressurized xenon gas is supplied to the system by a
specially constructed gas handling system. The majority of the gas
handling system is formed from 1/4 inch stainless steel piping with
Swagelok fittings mounted to a gas control panel. The same system was used to supply purified xenon to the NEXT-CRAB0 detector in Ref.~\cite{byrnes2023next}. 

Noble gases are supplied from cylinders with 99.999\% purity and circulated
 through hot~\footnote{SAES PS4-MT3-R-1} and cold~\footnote{SAES MicroTorr HP190-902F} zeolite getters to clean at $\sim$3 slpm
for several hours to remove oxygen, water and nitrogen contamination. Pumping action is provided by a hermetic, piston-driven gas pump with neoprene buffers~\footnote{PumpWorks PW2070} with pump-speed controlled by a variac on the power line. Experience with devices on the same gas system
show that this is sufficient to achieve part-per-billion levels
of oxygen and water impurity.  Gas pressure is monitored by several analog pressure gauges on the gas  panel and over-pressurization is prevented by a 400~psig relief on the panel, 250~psig relief on the vessel, and a 15~psig burst disk on the vacuum line. The temperature
inside the vessel is monitored by an internal thermocouple, and flow both
into and out of the vessel in standard liters per minute is monitored using
mass flow meters~\footnote{Omega FM1800}.

After running with argon, the gas is typically vented to
the room through a vent line, whereas due to its much higher cost xenon is recaptured into bottles
by cryogenic condensation with liquid nitrogen. A few psi of xenon is typically lost with each fill cycle due to incompleteness of this capture process. A manifold-like gas mixing
arrangement allows for use of multiple gases or mixtures, if needed.

\end{document}